\documentclass[a4paper,11pt]{article}

\usepackage[
    margin=0.8in
    ]{geometry}

\usepackage{amsmath,amssymb,amsthm,mathtools,mathrsfs,amsfonts,latexsym}
\usepackage[italicdiff]{physics}
\mathtoolsset{showonlyrefs,showmanualtags}
\numberwithin{equation}{section}
\allowdisplaybreaks[4]

\usepackage[
    backend=biber,
    bibstyle=phys,
    sorting=none,
    biblabel=brackets,
    citestyle=numeric-comp,
    pageranges=false,
    eprint=true,
    url=true,
    ]{biblatex}
\DeclareFieldFormat{url}{[\href{#1}{iNSPIRE}]}
\DeclareSourcemap{
	\maps[datatype=bibtex]{
		\map[overwrite=true]{
			\step[fieldsource=shortjournal]
			\step[fieldset=journal, origfieldval]
		}
        }
	\maps[datatype=bibtex]{
		\map[overwrite=true]{
			\step[fieldsource=shortbooktitle]
			\step[fieldset=booktitle, origfieldval]
		}
	}
}
\AtEveryBibitem{
    \iffieldequalstr{type}{NoiNSPIRE}{
        \DeclareFieldFormat{url}{[\href{#1}{Link}]}
        }{}
    \clearfield{note}
    \clearfield{type}
    \clearname{editor}
    \clearlist{location}
}
\bibliography{reference,fast-forward_Masuda}

\usepackage{hyperref}
\hypersetup{
    colorlinks=true,
    linktocpage=true,
    linkcolor=blue,
    urlcolor=blue,
    citecolor=blue,
    filecolor=blue,
    bookmarksopen=true,
    }

\usepackage{authblk}
\usepackage{orcidlink}

\title{Fast-forward scaling theory for quantum dynamics on curved space-time}
\author{Yuji Ando\thanks{E-mail: \href{mailto:y.andou@aist.go.jp}{y.andou@aist.go.jp}}}
\author{Shumpei Masuda\thanks{E-mail: \href{mailto:shumpei.masuda@aist.go.jp}{shumpei.masuda@aist.go.jp}}}
\affil{
Global Research and Development Center for Business by Quantum-AI Technology (G-QuAT),

National Institute of Advanced Industrial Science and Technology (AIST),

1-1-1 Umezono, Tsukuba, Ibaraki 305-8568, Japan}
\date{}

\begin{document}
\maketitle
\begin{abstract}
    Scaling properties inherent in quantum dynamics have been studied for various systems in terms of acceleration, deceleration and time reversing. We show a scaling property of quantum dynamics on curved space-time where gravity plays an essential role. We derive system parameters which realize speed-controlled dynamics. Moreover, we consider spatial scaling and derive system parameters which realize spatially-scaled quantum dynamics.
\end{abstract}
\thispagestyle{empty}
\newpage
\setcounter{page}{1}
\tableofcontents
\section{Introduction}
Speed has been a great interest of humans. The pursuit of speed has driven technological development in various fields. Motivated by the necessity of speed in quantum technology to avoid decoherence and by scientific curiosity, researchers have studied how to control the speed of quantum dynamics \cite{Masuda:2016,Masuda:2022nyu}. Previous work \cite{Masuda:2008} derived system parameters which realize speed-controlled dynamics of quantum systems, i.e., accelerated, decelerated and time-reversed dynamics. The theory called fast-forward scaling theory (FFST) reveals the scaling property in quantum dynamics, that is, a relation between dynamics with different speed. FFST was studied for various systems including charged particles \cite{MasudaPRA:2011}, qubits \cite{Masuda:2022pct}, discrete systems \cite{MasudaRice2014,Takahashi2014,Hatomura:2022huk}, and a many body system \cite{MASUDA2012}.

Finding a viable quantum evolution is time and resource consuming in general since it requires a numerical integration of an equation of motion such as the Schr\"{o}dinger equation. FFST can reduce the effort because once we find a viable quantum dynamics as a reference, the theory allows us to obtain infinite number of dynamics with different speed and corresponding system parameters without integration of the equation of motion. Thus, FFST was found useful for the parameter optimization for state preparation which aims at generating a desired state at a predetermined time \cite{Masuda:2016,Masuda:2022nyu}.

FFST was extended to accelerate quantum adiabatic dynamics which are ultimately slow dynamics \cite{Masuda:2010}. The theory can be categorized to shortcuts to adiabaticity (STA), a group of protocols which realize the desired end state of adiabatic dynamics in a short time \cite{Masuda:2016,Torrontegui:2013sge,delCampo:2019,Guery-Odelin:2019fcm,Hatomura:2023ijg}. FFST for quantum adiabatic dynamics has been studied in various systems such as cold atoms \cite{Masuda:2010,MugaJPB2009,TorronteguiPRA2011,MugaJPB2010,Xi2011,SchaffNJP2011,Torrontegui2012,TorronteguiNJP2012,MasudaPRL2014,TakahashiPRA2015,MartinezGaraot2016,JingJun2021}, charged particles \cite{MasudaPRA:2011,Kiely2015,MasudaJPCB2015,An:2016fkl}, many-body systems \cite{MASUDA2012}, spin systems \cite{Setiawan:2017,Setiawan:2019}, discrete systems \cite{MasudaRice2014,Takahashi2014,TakahashiPRA2015,Hatomura:2022huk}. Acceleration of classical adiabatic dynamics \cite{JarzynskiPRE2017} and the role of FFST in non-equilibrium statistical mechanics \cite{Babajanova2018,Nakamura2020} have been the subjects of research. Various approaches for fast-forwarding were developed to extend the applicability of the theory \cite{Patra2017,Patra2021,Masuda:2022pct}. Moreover, FFST has been studied for the Dirac equation—a representative relativistic quantum equation—in (1+1) and (2+1) dimensions \cite{Deffner:2016,Sugihakim:2021}.

Although the scaling property has been studied in various systems, it remains largely unexplored in quantum relativistic dynamics. Their previous works are not FFST in (3+1)-dimension. Thus, it is not trivial to apply their results to FFST in (3+1)-dimension and other relativistic quantum equations such as Klein-Gordon equation. The main objective of this work is to develop a deeper understanding of FFST in relativistic quantum dynamics and to obtain results that are broadly applicable. In relativistic theories, however, time and space are treated on equal footing due to Lorentz symmetry, and arbitrary rescaling of time alone is generally not allowed. This fundamental constraint prevents a straightforward application of non-relativistic fast-forward techniques, which often rely on asymmetric time reparameterization. To this end, we formulate FFST within the framework of general relativity rather than special relativity. In general relativity, spacetime is not necessarily flat but is determined by gravity, which is mathematically described by the metric tensor. This allows the equations of general relativity to remain covariant under coordinate transformations and facilitates the explicit construction of a combination of wave-function and metric that realizes speed-controlled dynamics. Additionally, we generalize the framework to include spatial scaling. While we focus on the Klein–Gordon equation as a representative example, our results are independent of the space-time dimension and can be extended to other relativistic quantum equations, in contrast to previous studies.

Importantly, our approach should not be misunderstood as a mere coordinate transformation of a known system. Instead, we explicitly construct a new spacetime geometry in which a speed-rescaled wavefunction becomes a genuine solution of the relativistic field equation, thereby embedding the desired dynamics into the curvature of spacetime itself.

This paper is organized as follows. In Sec \ref{Sec:review}, we briefly review FFST for non-relativistic quantum dynamics to illustrate the idea. In Sec \ref{Klein-Gordon equation}, we develop FFST for curved space-time using the Klein-Gordon equation. We conclude this paper in Sec \ref{summary}.

\section{Fast-forward scaling theory for non-relativistic quantum dynamics\label{Sec:review}}
Before we develop FFST for curved space-time, we illustrate the idea of FFST by briefly reviewing the theory for non-relativistic quantum dynamics \cite{Masuda:2022nyu}. We consider a particle with mass $m$ under a potential $V(t,\vb*{x})$. A viable quantum dynamics is represented by a wave function $\psi(t,\vb*{x})$ which is a solution of the Schr\"{o}dinger equation
\begin{align}
    i\hbar\pdv{t}\psi(t,\vb*{x})=\qty(-\frac{\hbar^2}{2m}\nabla^2+V(t,\vb*{x}))\psi(t,\vb*{x}).
\end{align}
We call this dynamics a reference dynamics of which speed is to be controlled.

A speed-controlled dynamics is represented as
\begin{align}
    \psi_\alpha(t,\vb*{x})=\psi(\Lambda(t),\vb*{x})\label{def:psi_alpha_7_16_24}
\end{align}
with
\begin{align}
    \Lambda(t)=\int_0^t\dd{t'}\alpha(t'),\label{def:Lambda_7_16_24}
\end{align}
where a real function $\alpha(t)$ is the speed-control factor. For example, the dynamics is accelerated for $\alpha>1$ and decelerated for $0<\alpha<1$. It is straightforwardly confirmed that modified mass and a potential defined by
\begin{align}
    m_\alpha&=\frac{m}{\alpha(t)},\\
    V_\alpha(t,\vb*{x})&=\alpha(t)V(t,\vb*{x}),\label{def:alpha_7_16_24}
\end{align}
can realize the speed-controlled dynamics \cite{Masuda:2008}. In other word, $\psi_\alpha(t,\vb*{x})$ is a solution of the Schr\"{o}dinger equation
\begin{align}
    i\hbar\pdv{t}\psi_\alpha(t,\vb*{x})=\qty(-\frac{\hbar^2}{2m_\alpha}\nabla^2+V_\alpha(t,\vb*{x}))\psi_\alpha(t,\vb*{x}).
\end{align}

The modulation of the mass expressed in Eq. \eqref{def:alpha_7_16_24} is undesirable from the viewpoint of control of the system. This difficulty is solved by using modulated wave function defined by
\begin{align}
    \psi_\mathrm{FF}(t,\vb*{x})=e^{if(t,\vb*{x})}\psi_\alpha(t,\vb*{x}),\label{def:FF_7_16_24}
\end{align}
where $f(t,\vb*{x})$ is real function and it is called additional phase. We assume that $\psi_\mathrm{FF}(t,\vb*{x})$ satisfies the Schr\"{o}dinger equation 
\begin{align}
    i\hbar\pdv{t}\psi_\mathrm{FF}(t,\vb*{x})=\qty(-\frac{\hbar^2}{2m}\nabla^2+V_\mathrm{FF}(t,\vb*{x}))\psi_\mathrm{FF}(t,\vb*{x}),\label{Sceq:FF_7_16_24}
\end{align}
where $V_\mathrm{FF}(t,\vb*{x})$ is real and it is called driving potential. By using Eqs. \eqref{def:psi_alpha_7_16_24}, \eqref{def:Lambda_7_16_24}, \eqref{def:alpha_7_16_24}, \eqref{def:FF_7_16_24}, and \eqref{Sceq:FF_7_16_24}, we can explicitly obtain the additional phase and the driving potential as \cite{Masuda:2022nyu}
\begin{align}
    f(t,\vb*{x})&=[\alpha(t)-1]\eta,\\ 
    V_\mathrm{FF}(t,\vb*{x})&=V(\Lambda(t),\vb*{x})-\hbar\pdv{\alpha(t)}{t}\eta-\hbar\frac{\alpha^2(t)-1}{\alpha(t)}\pdv{t}\eta\\
    &\qquad-\frac{\hbar^2}{2m}[\alpha^2(t)-1][\nabla\eta]^2,
\end{align}
where $\eta$ abbreviates $\eta(\Lambda(t),\vb*{x})$. Here, $\eta(t,\vb*{x})$ is the phase of the reference wave function, that is, the reference wave function is written as $\psi(t,\vb*{x})=\tilde{\psi}(t,\vb*{x})\exp[i\eta(t,\vb*{x})]$ with real functions $\tilde{\psi}(t,\vb*{x})$, $\eta(t,\vb*{x})$.

The above results show that once we obtain a single viable quantum dynamics we can obtain speed-controlled dynamics and corresponding driving potentials without integration of the Schr\"{o}dinger equation. In the following section, we present such a scaling property in quantum relativistic dynamics.

\section{Fast-forward scaling theory for curved space-time\label{Klein-Gordon equation}}
We develop FFST for relativistic quantum mechanics, particularly, a scalar field $\phi$. For later convenience, we consider scalar field on curved space-time. Namely, the scalar field couples to gravity $g_{\mu\nu}$ and it follows the Klein-Gordon equation on curved space-time:
\begin{align}
    0=\frac{1}{\sqrt{-g(t,\vb*{x})}}\partial_\mu\qty(\sqrt{-g(t,\vb*{x})}g^{\mu\nu}(t,\vb*{x})\partial_\nu\phi(t,\vb*{x}))-\qty(\frac{mc}{\hbar})^2\phi(t,\vb*{x}),\label{KG_9_12_24}
\end{align}
where $g_{\mu\nu}$ is a fixed background; $g$ is determinant of $g_{\mu\nu}$; $m$ is the mass of the particle. Also, we use the Einstein's summation convention\footnote{See textbooks \cite{Hawking:1973uf,Zee:2013dea} for reviews of general relativity.}
and the Greek indices $\mu,\nu$ run over $0,1,\dots,d$ where $d$ is the dimension of the space-time. In the following, we set $d$ to three without loss of generality.

A speed-controlled dynamics is given by $(\phi(\Lambda(t),\vb*{x}),g_{\mu\nu}(\Lambda(t),\vb*{x}))$ and they follow
\begin{align}
    0&=\frac{1}{\sqrt{-g(\Lambda(t),\vb*{x})}}\partial'_\mu\qty(\sqrt{-g(\Lambda(t),\vb*{x})}g^{\mu\nu}(\Lambda(t),\vb*{x})\partial'_\nu\phi(\Lambda(t),\vb*{x}))\\
    &\qquad\qquad-\qty(\frac{mc}{\hbar})^2\phi(\Lambda(t),\vb*{x}),\label{KG2_9_12_24}
\end{align}
where $\partial'_\mu,\partial'^\mu$ are the derivative with respect to $x'^\mu=(\Lambda(t),\vb*{x})$. We now aim at finding system parameters which generate the speed-controlled state. First, we rewrite the equation, originally formulated in terms of $(\Lambda(t),\vb*{x})$, using the original coordinates $(t,\vb*{x}))$. By applying the chain rule, the derivatives $\partial'_\mu$ and $\partial'^\mu$ expressed in terms of the original coordinates as
\begin{align}
    \partial'_\mu=\pdv{x^\nu}{x'^\mu}\partial_\nu\qc\partial'^\mu=\pdv{x_\nu}{x'_\mu}\partial^\nu.\label{derivative_9_10_24}
\end{align}
However, unlike in non-relativistic systems where time and space derivatives can often be treated independently, relativistic field equations couple temporal and spatial derivatives through the spacetime metric. As a result, a naive rescaling of the time coordinate leads to nontrivial modifications in the structure of the field equations, and one cannot simply carry over the techniques used in non-relativistic fast-forward scaling. See Appendix \ref{Intruduction of additional phase}.

Nevertheless, by working within the framework of general relativity, we can overcome this difficulty. In general relativity, the dynamics of a field are fully determined by its coupling to the spacetime geometry through the metric tensor. This geometric structure allows us to define the speed-controlled state $\phi_\mathrm{FF}(t,\vb*{x})$ and metric $(g_\mathrm{FF})_{\mu\nu}(t,\vb*{x})$.
\begin{align}
    \phi_\mathrm{FF}(t,\vb*{x})&\coloneqq\phi(\Lambda(t),\vb*{x})\label{phiFF_9_12_24}\\
    (g_\mathrm{FF})_{\mu\nu}(t,\vb*{x})&\coloneqq{\pdv{x'^\rho}{x^\mu}\pdv{x'^\sigma}{x^\nu}}g_{\rho\sigma}(\Lambda(t),\vb*{x})\label{def:gFF}
\end{align}
Using them and Eq. \eqref{derivative_9_10_24}, the first term of the right hand side of Eq. \eqref{KG2_9_12_24} is rewritten as
\begin{align}
    &\frac{1}{\sqrt{-g(\Lambda(t),\vb*{x})}}\partial'_\mu\qty(\sqrt{-g(\Lambda(t),\vb*{x})}g^{\mu\nu}(\Lambda(t),\vb*{x})\partial'_\nu\phi(\Lambda(t),\vb*{x}))\\
    &=\frac{1}{\sqrt{-g_\mathrm{FF}(t,\vb*{x})}}\partial_\mu\qty(\sqrt{-g_\mathrm{FF}(t,\vb*{x})}g_\mathrm{FF}^{\mu\nu}(t,\vb*{x})\partial_\nu\phi_\mathrm{FF}(t,\vb*{x})).\label{eq1_9_12_24}
\end{align}
Therefore, we obtain
\begin{align}
    0&=\frac{1}{\sqrt{-g_\mathrm{FF}(t,\vb*{x})}}\partial_\mu\qty(\sqrt{-g_\mathrm{FF}(t,\vb*{x})}g_\mathrm{FF}^{\mu\nu}(t,\vb*{x})\partial_\nu\phi_\mathrm{FF}(t,\vb*{x}))\\
    &\qquad\qquad-\qty(\frac{mc}{\hbar})^2\phi_\mathrm{FF}(t,\vb*{x}).
\end{align}
General covariance guarantees that the pair $(\phi_\mathrm{FF}(t,\vb*{x}),(g_\mathrm{FF})_{\mu\nu}(t,\vb*{x}))$, expressed in the original coordinates $(t,\vb*{x}))$, is physically equivalent to $(\phi(\Lambda(t),\vb*{x}),g_{\mu\nu}(\Lambda(t),\vb*{x}))$, which is written in the rescaled coordinates $(\Lambda(t),\vb*{x})$. Physically, this result shows that the gravity $(g_\mathrm{FF})_{\mu\nu}$ can realize the speed controlled dynamics.

Now, a few comments are in order. The gravity $(g_\mathrm{FF})_{\mu\nu}$ depends only on speed-control factor $\alpha$ and is independent of the reference dynamics. If we set $g_{\mu\nu}$ to the Minkowski metric $\eta_{\mu\nu}=\mathrm{diag}[-1,1,1,1]$, namely $g_{\mu\nu}=\eta_{\mu\nu}$, we have $(g_\mathrm{FF})_{\mu\nu}=\mathrm{diag}[-\alpha^2(t),1,1,1]$. The mass of the particle does not have to be changed in contrast to Eq. \eqref{def:alpha_7_16_24}. The speed-controlled state with modulated phase $\phi_\mathrm{FF}(t,\vb*{x})= e^{if(t,\vb*{x})}\phi_\alpha(t,\vb*{x})$ cannot be realized with fixed gravity as explained in Appendix \ref{Intruduction of additional phase}.

\subsection{Newton potentials}
So far, we showed that a speed-controlled dynamics is realized under a gravitational field in Eq. \eqref{def:gFF}. Now, we examine Newton potentials for the reference and speed-controlled dynamics by using the Newtonian approximation which is valid when the following conditions are satisfied.
\begin{enumerate}
    \item Static spacetime : $\partial_0g_{\mu\nu}=g_{0i}=0$
    \item Weak-field approximation : $\abs{g_{\mu\nu}-\eta_{\mu\nu}}\ll1$
    \item Non-relativistic limit : $v^2\ll c^2$
\end{enumerate}
The Newton potential for the reference dynamics $\varphi$ is related to $g_{00}$ and $\eta_{00}$ as\footnote{See Chapter V.4 in textbook \cite{Zee:2013dea} for more details.}
\begin{align}
    \partial_\mu\varphi=-\frac{c^2}{2}\partial_\mu(g_{00}-\eta_{00}),\label{phi_9_13_24}
\end{align}
while the Newton potential for the speed-controlled dynamics $\varphi_\mathrm{FF}$ satisfies
\begin{align}
    \partial_\mu\varphi_\mathrm{FF}=-\frac{c^2}{2}\partial_\mu((g_\mathrm{FF})_{00}-\eta_{00}).\label{phiFF_9_13_24}
\end{align}
Because $(g_\mathrm{FF})_{\mu\nu}$ is defined as Eq. \eqref{def:gFF}, we have
\begin{align}
    (g_\mathrm{FF})_{00}=\alpha^2g_{00}.\label{gcom_9_13_24}
\end{align}
Using Eqs. \eqref{phi_9_13_24}, \eqref{phiFF_9_13_24}, and \eqref{gcom_9_13_24}, we obtain
\begin{align}
    \partial_\mu\varphi_\mathrm{FF}=\alpha^2\partial_\mu\varphi.
\end{align}
This is consistent with the result for a classical dynamics (see Appendix \ref{Classical dynamics under gravity}).

\subsection{Spatial scaling}
This theory can be extended to spatial scaling. We consider spatially-scaled dynamics
\begin{align}
    \phi_\mathrm{SS}(t,\vb*{x})\coloneqq\phi(t,\Lambda^{(x)}(x),y,z),
\end{align}
where 
\begin{align}
    \Lambda^{(x)}(x)=\int_0^x\dd{x'}\alpha^{(x)}(x').
\end{align}
Here, $\phi(t,\vb*{x})$ is a solution of the Klein-Gordon equation with the Minkowski metric. The scalar field of the scaled dynamics is compressed (expanded) along the $x$-axis for $\alpha>1$ ($\alpha<1$) compared to the reference one. In an analogous manner, we can derive the metric which generates $\phi_\mathrm{SS}$ as
\begin{align}
    (g_\mathrm{SS})_{\mu\nu}=\mathrm{diag}[-1,(\alpha^{(x)})^2,1,1].
\end{align}

We further generalize the theory by considering the scaling of time and space
\begin{align}
    \phi_\mathrm{SS}(t,\vb*{x})\coloneqq\phi_0(\Lambda^{(t)}(t),\Lambda^{(x)}(x),\Lambda^{(y)}(y),\Lambda^{(z)}(z)),
\end{align}
where $\Lambda^{(j)}$ is defined by 
\begin{align}
    \Lambda^{(j)}(j)=\int_0^j\dd{j'}\alpha^{(j)}(j').
\end{align}
for $j=\{t,x,y,z\}$. The metric which generates the scaled dynamics is given as
\begin{align}
    (g_\mathrm{SS})_{\mu\nu}=\mathrm{diag}[-(\alpha^{(t)})^2,(\alpha^{(x)})^2,(\alpha^{(y)})^2,(\alpha^{(z)})^2].
\end{align}

\section{Summary and discussion\label{summary}}
We have developed FFST on curved space-time presenting a scaling property in relativistic quantum dynamics. We have derived metrics which generate speed-controlled dynamics. Moreover, we have extended the theory to spatial scaling showing system parameters which generate spatially-scaled dynamics.

Our contribution shows that, rather than working in a transformed coordinate system with a fixed metric, one can retain the original coordinates and instead modify the metric to realize the desired time-rescaled dynamics. This approach preserves the clarity and structure of the original equation while enabling speed control via geometry. Also, while we do not claim that spacetime metrics can be directly engineered in practice, it is important to note that, according to general relativity, modifying the gravitational field is locally equivalent to considering an accelerated frame. From this perspective, the geometric formulation we provide may be applicable to systems where acceleration can be externally controlled, such as in quantum systems undergoing non-inertial motion. This opens a potential avenue for exploring speed-controlled quantum dynamics through effectively curved space-times realized by accelerated reference frames, which are within the reach of laboratory conditions. Therefore, this is both mathematically precise and potentially useful for physical applications.

In this paper, we focused on the free scalar field following the Klein-Gordon equation, but our results are more broadly applicable. The reason is that any equation of motion on curved space-time is given by requiring covariance under general coordinate transformation. For example, the equation of motion in an interacting scalar field theory on curved space-time differs from Eq. \eqref{KG_9_12_24} but, by construction, it has to be covariant under general coordinate transformation. As a result, even if there are interaction terms, $(g_\mathrm{FF})_{\mu\nu}$ given by Eq. \eqref{def:gFF} can realize the speed controlled dynamics.

Moreover, the same results can be obtained for not only scalar field but also spinor field, vector field and so on. For example, if a vector field $A_\mu$ and $g_{\mu\nu}$ satisfy an equation of motion on curved space-time, the covariance under general coordinate transformation guarantees that $(A_\mathrm{FF})_{\mu}$ and $(g_\mathrm{FF})_{\mu\nu}$ given by
\begin{align}
    (g_\mathrm{FF})_{\mu\nu}(t,\vb*{x})&\coloneqq{\pdv{x'^\rho}{x^\mu}\pdv{x'^\sigma}{x^\nu}}g_{\rho\sigma}(\Lambda(t),\vb*{x}),\\
    (A_\mathrm{FF})_\mu(t,\vb*{x})&\coloneqq{\pdv{x'^\rho}{x^\mu}}A_{\rho}(\Lambda(t),\vb*{x}),
\end{align}
satisfy the equation of motion. An important point is that the definition of $(g_\mathrm{FF})_{\mu\nu}$ is the same as Eq. \eqref{def:gFF}. Namely, the definition of $(g_\mathrm{FF})_{\mu\nu}$ which can realize the speed-controlled dynamics does not depend on whether we consider scalar field or other fields.

By considering the covariance on curved space-time, we can simply obtain the speed-controlled state $\phi_\mathrm{FF}$ and $(g_\mathrm{FF})_{\mu\nu}$. Our result highlights the usefulness of the covariance under general coordinate transformation for FFST. Hence, in this paper, while we considered relativistic quantum mechanics on curved space-time, it may also be useful to consider non-relativistic quantum mechanics on curved space-time. Such theory can be constructed using Newton–Cartan geometry \cite{Baiguera:2023fus} and it would be intriguing to examine the relation between the derivation using Newton–Cartan geometry and the previous work of FFST for non-relativistic quantum mechanics.
\appendix
\section{Speed-controlled state with modulated phase\label{Intruduction of additional phase}}
We show that the modified speed-controlled state defined by
\begin{align}
    \phi_\mathrm{FF}(t,\vb*{x)}= e^{if(t,\vb*{x})}\phi(\Lambda(t),\vb*{x})
\end{align}
is not realizable with the same gravity as the reference one in general. The state is realizable if there exist $\phi_\mathrm{FF}(t,\vb*{x})$ which satisfies the Klein-Gordon equation
\begin{align}
    0=\frac{1}{\sqrt{-g(t,\vb*{x})}}\partial_\mu\qty(\sqrt{-g(t,\vb*{x})}g^{\mu\nu}(t,\vb*{x})\partial_\nu\phi_\mathrm{FF}(t,\vb*{x}))-\qty(\frac{mc}{\hbar})^2\phi_\mathrm{FF}(t,\vb*{x}),\label{KGFF_9_12_24}
\end{align}
for given $m$, $g^{\mu\nu}$ and $\phi$ which satisfies Eq. \eqref{KG_9_12_24}. However, there is no such $f$ in general. By separating Eq. \eqref{KGFF_9_12_24} into the real and imaginary parts, it is viewed as two simultaneous differential equations. Because we have a single free variable $f(t,\vb*{x})$ at each $(t,\vb*{x})$, we cannot satisfy the simultaneous equations.

In contrast to the case of the Klein-Gordon equation, the modified speed-controlled state can be realized for Schr\"{o}dinger equation as explained in Sec \ref{Sec:review}. This is because that we have two parameters, i.e., the additional phase and the potential in the Schr\"{o}dinger case.

\section{Classical dynamics under gravity\label{Classical dynamics under gravity}}
We consider a classical dynamics of a particle under the gravity. The dynamics is governed by the equation of motion
\begin{align}
    \dv[2]{x}{t}=-g,\label{EOMC_9_5_24}
\end{align}
where $g$ is the gravitational acceleration. Suppose that $x(t)$ is a solution of Eq. \eqref{EOMC_9_5_24}. We define a speed-controlled dynamics as 
\begin{align}
    x_\alpha(t)=x(\Lambda(t)),\label{x_alpha_9_5_24}
\end{align}
with $\Lambda(t)$ defined by Eq. \eqref{def:Lambda_7_16_24}. Using Eqs. \eqref{x_alpha_9_5_24} and \eqref{def:Lambda_7_16_24}, we obtain 
\begin{align}
    \dv[2]{x_\alpha(t)}{t}=-\alpha^2(t)g.\label{EOMC_9_5_24_2}
\end{align}

Now, we consider a particle under the gravity with the gravitational acceleration $g_\alpha$, and assume that $x_\alpha(t)$ is a viable dynamics of the particle. Then, we have 
\begin{align}
    \dv[2]{x_\alpha(t)}{t}=-g_\alpha.\label{EOMC_9_5_24_3}
\end{align}
By comparing Eq. \eqref{EOMC_9_5_24_2} with Eq. \eqref{EOMC_9_5_24_3}, we obtain the gravitational acceleration for the speed-controlled dynamics as
\begin{align}
    g_\alpha=\alpha^2(t)g.
\end{align}
\printbibliography
\end{document}